\documentclass[aps,prb,notitlepage]{revtex4-1}
\usepackage{graphicx}
\usepackage{graphics}
\usepackage{amssymb,amsmath}

\begin{document}

\title{Self-Assembled Nanostructure Formation in Chemical Vapor Deposition.}
\author{Daniel Walgraef}
\email[E-mail me at:]{ dwalgraef@ifisc.uib-csic.es}
\affiliation{
IFISC (CSIC-UIB), Instituto de F\'isica Interdisciplinar y Sistemas Complejos, E-07122 Palma de Mallorca, Spain (http://www.ifisc.uib.es).}

\begin{abstract}
When thin films are grown on a substrate by chemical vapor deposition, the evolution of the first deposited layers may be described, on mesoscopic scales, by dynamical models of the reaction-diffusion type. For monoatomic layers, such models describe the evolution of atomic coverage due to the combined effect of reaction terms representing adsorption-desorption and chemical processes and nonlinear diffusion terms which are of the Cahn-Hilliard type. This combination may lead, below a critical temperature, to the instability of uniform deposited layers. This instability triggers the formation of nanostructures corresponding to regular spatial variations of substrate coverage. Patterns wavelengths and symmetries are selected by dynamical variables and not by variational arguments. According to the balance between reaction- and diffusion-induced nonlinearities, a succession of self-assembled structures including hexagonal arrays of dots, stripes and localized structures of various types may be obtained. These structures may initiate different types of growth mechanisms, including Volmer-Weber and Frank-Van der Merwe type of growth. The relevance of this approach to the study of different types of deposited monolayers is discussed.\\

Keywords: Thin film growth, nonlinear deposition rate, nonlinear diffusion, nanostructure formation, localized dots.

PACS: 68.55.-a, 62.23.St, 81.15.Aa, 81.15.Gh
\end{abstract}
\maketitle

\section{Introduction.}
 Overwhelming progress has recently been made in the area of nanoscale science and technology. For example, the formation of self-assembled nanostructures in deposited layers on solid surfaces has become the subject of intense research activity, due to its fundamental and technological relevance. Due to the ever growing technological importance of on-demand tailored nanomaterials, it is of capital economic and scientific interest to understand and master their formation and growth, since growth mechanisms usually determine most of their properties and textures. Furthermore, the ability to probe structure-property relationships on an appropriate length scale is a primary driver of progress in nanotechnology. As a result, the modelling of nanostructure formation and properties is one of the most active research areas in materials science.

For thin film growth, for example, numerous computer simulation methods have been developed since the 1970's. These investigations revealed detailed information on growing films, such as island shapes, step formation and surface roughening \cite{atfilmgrowth1,atfilmgrowth2}. Unfortunately, in molecular dynamics and Monte-Carlo approaches, the computational time required to simulate thin film growth under realistic deposition rates are often excessive. Effectively, a significant constraint of these methods is that, due to their atomistic nature, they can deal only with small systems, with linear dimension under the micron. Other approaches, based on discrete lattices, are able to simulate thin film growth under realistic deposition rates even in three dimensions. Nevertheless, attempts to bridge molecular dynamics and Monte Carlo method succeeded in simulating small polycrystalline films \cite{hanchen}. It is why continuum approaches are particularly interesting, since they are capable of simulating growth of thin films of larger dimensions \cite{cale}. In these methods, the film surface is represented by a series of mesh points that move according to materials flux exchanged between neighboring regions. The drawback of continuum simulations is that they provide information about the film surface only. Textures and microstructures are thus ignored in these descriptions. On the other hand, continuous models, based on rate equations of the PDE type, have also been proposed to describe mesoscopic scales, with the hope to improve the understanding of deposition processes and film surface evolution at scales inaccessible by both traditional equipment (macroscopic) and feature scales (microscopic) models \cite{gaonix}.

Up to now, these models had limited predictive capability, because of a rough description of kinetic processes, such as atomic diffusion or deposition.  Nevertheless, such capability could be greatly enhanced in the framework of Multiscale Modelling. Effectively, the concept of Multiscale Materials Modelling has been recently introduced to bridge the gaps between atomistic and continuum methods, and to link them in a consistent way \cite{MMM}. The aim is to obtain a reliable description of materials behavior, from microscopic to macroscopic scales. This program should be realized by coupling models for different length scales. The results from smaller scales are fed to larger scales, with appropriate mesh redefinition, and the results from larger scales are being fed back to the smaller ones, in a back and forth process hopefully ending in quantitatively reliable solution. In the case of thin film growth, if information from each scale is transferred correctly to the other scales, one would expect to be able to follow the evolution of film textures, surface topography, the effect of microstructures on local deposition rates, etc... . In this framework, the interest of mesoscopic continuous models would be to link micro- and macroscales, and provide not only qualitative, but also quantitative descriptions of thin film growth and nanostructure formation.

In this framework, continuous mesoscopic models have already been proposed to describe the spontaneous ordering of nanostructures or quantum dots in multicomponent epilayers on a substrate \cite{desai-98,suo-00b}. Ordering result from the balance between coarsening effects induced by the spinodal decomposition of the solid solution which forms the film, and phase refining induced by concentration-dependent surface stresses. Nevertheless we think that the instability of the deposited layer and the formation of self-organized nanophases should have a dynamical origin. Furthermore it should occur either in monoatomic layers and binary epilayers, as suggested in \cite{mono_1,mono_2} and \cite{binary_1,binary_2}. It is why we proposed, a decade ago, a reaction-diffusion model to describe, at the nanoscale, structure formation and texture evolution in an adsorbed monolayer on a substrate \cite{walgraef_02a,walgraef_02b}. With this model, we have shown that, even in monocomponent films, the competition between atomic deposition and the underlying instability of an adsorbed atomic layer may stabilize nanoscale spatial patterns, already in the first deposited layers. These patterns correspond to regular spatial distributions of high and low coverage domains, which may induce corresponding distributions of grains with different orientations or symmetries, and serve as templates for the later stages of film textures evolution. Patterns with different symmetries may be selected, according to the relative values of experimental parameters such as deposition rate, substrate temperature and atomic mobility. In systems with isotropic diffusion, successive transitions between hexagonal arrays of dense-on-dilute spots, stripes and hexagonal arrays of dilute-on-dense spots are predicted for increasing coverage, or concentration. It is worth noting that, in this approach, pattern selection has a dynamical origin, in contrast with the previous methods, which are based on the minimization of a free energy \cite{suo-00a,suo-00b}, or to dynamical descriptions of nanostructures induced by spinodal decomposition in multicomponent films \cite{desai-97,desai-98}. This model has been extended to include long-range interactions between deposited atom clusters mediated by the substrate \cite{hu}. In this case cluster-cluster interactions slightly favor stripes. The length scale of natural selforganized structures has been found to be in the tens of nanometers range. Imposition of a substrate periodic strain field by subsurface interfacial dislocations has also been considered and has been shown to dramatically change the self-organized pattern and its length scale. Qualitative agreements between model predictions and experimental observations on self-organized Ge quantum dots on Si substrate has been obtained \cite{kim1,kim2}. This model has also been rederived, including more general desorption processes, and has been shown to generate highly nonlinear structures, corresponding to regular arrays of low coverage holes in high coverage background, or to regular arrays of dots in a low coverage background, as well as stable isolated localized structures \cite{marcel}.

Up to know the simplest deposition processes have been considered, which depend linearly on the atomic coverage of the growing film and may correspond to sputtering or laser assisted deposition. In this case, the nonlinear part of the dynamics comes from the atomic diffusion in the first deposited layers only. However, in chemical vapor deposition, the reaction part of the dynamics may also be nonlinear. Hence, the competition between nonlinearities arising from reaction and diffusion terms could affect pattern formation, selection and stability in the deposited layer. It is why my aim in this paper is to present a few mesoscopic models dedicated to the modelling of thin films deposited on a substrate through chemical vapor deposition and to analyze the effect of more complex adsorption-desorption mechanisms on their stability and self-assembled nanostructure formation.

The paper is organized as follows. First, the dynamics of a deposited atomic layer on a substrate is reviewed and a reaction-diffusion model appropriate for chemical vapor deposition processes is derived in section \ref{dynmodel}. The stability of uniform deposited layers is analyzed in section \ref{linear stability}. Weakly nonlinear dynamics is performed close to instability and resulting pattern formation and selection beyond instability is analyzed for different types of nonlinearities and diffusion coefficients in section \ref{nonlinear}. The possibility of experimental realisation of the proposed instability mechanism is discussed in section \ref{discussion}. Finally, conclusions and perspectives for future work are presented in section \ref{conclusions}.

\section{The Dynamics of a Deposited Layer on a Substrate.}\label{dynmodel}

As already discussed in previous publications, the evolution of a monoatomic layer, deposited on a substrate, may be described by a continuous dynamical model of the reaction-diffusion type  \cite{walgraef_02a,walgraef_02b}. Relevant examples of such systems are $Al$ or $Cu$ layers deposited on $Si$ substrates, or $SiO_2$ and $TiN$ layers deposited on $Ti$ or $Al$ substrates. In such cases, the dynamics is governed by atomic adsorption, desorption and reaction on the substrate, but also by diffusion or transport. For sufficiently small lattice misfit between film and substrate, elasticity and stress effects may be neglected, and the film evolution may be described by atomic coverage dynamics only. On mesoscopic scales, the corresponding kinetic equation has been derived in the framework of chemical kinetics, and has the following structure \cite{walgraef_02a}:
\begin{equation}\label{GenNLDiffusion}
\partial_t c = R(c) - \vec\nabla\vec J
\end{equation}
where $c = c(\vec r,t)$ is the local atomic coverage of the substrate, which is defined as the average occupancy number, or average atom number per lattice site.  $R(c)$ represents reaction terms and $\vec J$ is the atomic current in the deposited layer, which is assumed to remain in local thermodynamical equilibrium.

\subsection{The dynamics of chemical vapor deposition.}\label{CVD}

In chemical vapor deposition (CVD), typical procedures consist in the deposition of precursor molecules, which contain the active species of the growing film, on the substrate. These molecules dissociate through chemical reactions and liberate the active species atoms which remain adsorbed on the substrate to form the film. Part of them may recombine and desorb. Precursor dissociation may be autocatalytic, and the chemical scheme of the process is of the type:
\begin{eqnarray}
P_g + n* (+ mC)\to P (+ mC)\nonumber \\
P + pC\to qC \nonumber \\
rC\to B \nonumber \\
B\to B_g+ r*
\end{eqnarray}
where $*$ represents vacant lattice sites, and the subscript $_g$ represents gas phase concentration. $P$ represents precursor molecules. $C$ is the active species which forms the film and which also recombines in desorbing molecules $B$. For large Knudsen number, the precursor transport is ballistic, and one may neglect flow transport processes of the precursor, its interaction with the surface being the dominant process. $C$ is then the only diffusing species, and the kinetic equations for the corresponding coverages may be written as:
\begin{eqnarray}\label{CVDdyn}
\partial_t P&=& \alpha (1-C)^nC^m - \rho PC^p \nonumber \\
\partial_t C&=& \rho PC^p - \beta C^r - \vec\nabla\vec J \nonumber \\
\partial_t B&=& \beta C^r - \kappa B
\end{eqnarray}
where $n$ represents the number of lattice sites required for the adsorption of a $P$ molecule, and $m$ represents the autocatalytic nature of the process.

Non trivial uniform steady states are given by $B_0 = \frac{\beta}{\kappa} C_0^r$, $P_0=\frac{\beta}{\rho} C_0^{r-p}$ and $(1-C_0)^nC_0^{m-r} = \frac{\beta}{\alpha}$ (with $n\ge 1$ and $r\ge 2$ in non trivial cases). For $m < r$, there is always a unique non trivial solution. For $m\ge r$, two nontrivial solution exist, if $\frac{\beta}{\alpha} < \frac{n^n(m-r)^{m-r}}{(m+n-r)^{m+n-r}}$. In the absence of transport, one of these solutions is stable, the other one is unstable. The stable solution is such that $\frac{m-r}{m+n-r}<c_0<1$. Furthermore, the adiabatic elimination of  adsorbed and desorbed species in (\ref{CVDdyn}) leads to the following kinetic equation for the active species coverage:
\begin{equation}\label{coveragedynCVD}
\partial_t c = \alpha (1-c)^nc^m - \beta c^r - \vec\nabla\vec J(c)
\end{equation}

\begin{itemize}

\item For $m < r$, the steady state condition writes
\begin{equation}\label{steady1}
(\alpha (1-c)^n - \beta c^{r-m})c^{m}=0
\end{equation}
and there is one trivial $c_0=0$ and one nontrivial $c_0\ne 0 $ solution. The evolution of perturbations $\sigma$ of the trivial solution being given by
\begin{equation}\label{stab1a}
\partial_t \sigma = (\alpha (1-\sigma )^n - \beta \sigma^{r-m})\sigma^{m}\simeq \alpha  \sigma^{m}
\end{equation}
it turns out that this solution is unstable. On the other hand, the linear evolution of perturbations of the nontrivial solution being
\begin{equation}\label{stab1b}
\partial_t \sigma = -n(\alpha (1- c_0 )^{n-1} + (r-m)\beta c_0^{r-m-1})c_0^{m}\sigma <0
\end{equation}
this steady state is stable.

\item For $m= r$, besides a trivial steady state, there is one nontrivial steady state if $\beta < \alpha$. In this case, the trivial steady state is unstable while the nontrivial one is stable.

\item For $m> r$, besides the trivial steady state, there a two nontrivial ones given by
\begin{equation}\label{steady2}
\alpha (1-c_0)^nc_0^{m-r} - \beta = 0
\end{equation}
The stability of the trivial solution results from
\begin{equation}\label{stab3a}
\partial_t \sigma = (\alpha (1-\sigma )^n\sigma^{m-r} - \beta )\sigma^{r}\simeq -\beta  \sigma^{r}<0
\end{equation}
which shows that it is stable. The linear stability of the nontrivial steady states results from
\begin{equation}\label{stab3b}
\partial_t \sigma =  \beta (n+m-r)\frac{c_{max}-c_0}{c_0(1-c_0)}\sigma
\end{equation}
and the state such that $c_0 < c_{max}$ is unstable while the state such that $c_0 > c_{max}$ is stable.

\end{itemize}

\subsection{Transport.}\label{transport}

Since it has been assumed that the active species atoms diffuse on the surface, one may use the formulation of \cite{walgraef_02b,marcel} where diffusion is governed by a spatially varying chemical potential. Since the chemical potential is the functional derivative of the free energy, one has:
\begin{equation}
\vec J = -\, L \, \vec\nabla \frac{\delta{\cal F}}{\delta c}
\end{equation}
where ${\cal F}$ is the free energy of the adsorbed layer formed by the active species $C$. In the mean field approximation, an explicit expression for the current may be obtained, which reads \cite{walgraef_02b,marcel}:

\begin{equation}\label{adsenergy}
{\cal F} = \int_S d{\mathbf r}\biggl( k_BT f({\mathbf r})- \frac{1}{2}\epsilon_0c({\mathbf r})^2 + \frac{1}{2}  \xi_0^2\vert\nabla c ({\mathbf r})\vert^2  \biggr)
\end{equation}
where $c ({\mathbf r})$ is the local coverage, $f({\mathbf r})= (1-c ({\mathbf r}))\ln (1-c ({\mathbf r})) +  c ({\mathbf r})\ln c ({\mathbf r})-\epsilon_0 c $. For nearest neighbor attractive interactions between deposited atom, $\epsilon_0=\gamma\epsilon$ and $\xi_0^2 = \gamma\epsilon l^2$, where $\gamma$ is the lattice coordination number, $\epsilon$ is the pair interaction energy and $l$ the lattice constant. Interaction between the substrate and adsorbed particles may be introduced through an extra contribution equal to $\epsilon_s c ({\mathbf r})$ in the local free energy. The local chemical potential may then be easily obtained from the resulting free energy, and writes:

\begin{equation}\label{chempot}
\mu ({\mathbf r}) = \frac{\delta{\cal F}} {\delta c} = \epsilon_s - \epsilon_0c ({\mathbf r}) + k_BT\ln \frac{c ({\mathbf r})}{ 1-c ({\mathbf r})} + \xi_0^2\nabla^2c ({\mathbf r})
\end{equation}
This is the equation of state of the system which defines the coverage as a function of temperature, interaction energies, etc. Let me recall that thermodynamic stability of a state $c ({\mathbf r})$ requires

\begin{equation}
\frac{\partial^2 f}{\partial c^2} = -\epsilon_0 + k_BT \frac{1}{c (1-c )} > 0
\end{equation}
and this condition is always satisfied for $\epsilon_0 < 4k_BT$, or $T>T_c$, with $\epsilon_0 = 4k_BT_c$. However, for $T<T_c$, states in the range
\begin{equation}
\frac{1}{2} [1 - \sqrt{1 - \frac{T}{T_c}}]  < c < \frac{1}{2} [1 + \sqrt{1 - \frac{T}{T_c}}]
\end{equation}
are thermodynamically unstable. Hence, in this range and without reacting terms, single homogeneous phases are unstable, and, due to coarsening, the system separates into two distinct phases, one with low coverage ($c < \frac 12$), the other one with high coverage ($c > \frac 12$). $T=T_c$ defines the critical point below which phase separation occurs in the adsorbed layer. The corresponding critical coverage and  chemical potential are $c_c = \frac 12$ and $\mu_c = \epsilon_s - 2k_BT_c$, respectively.

Furthermore, $L =\frac{D}{k_BT}$, where $D$ is the surface diffusion coefficient. For Fick diffusion, it may be considered as a constant, although, according to the dynamical processes involved in atomic displacements, it may also be coverage dependent. For example, expressions such as $D = D_0 c(1-c)$ may be more appropriated for hopping types of motion, with $D_0 \propto \exp -(\Omega /T)$, where $\Omega$ is related to the activation energy for atomic jumps.

\section{The stability of uniform deposited layers.}\label{linear stability}

Since the presence of reacting terms is expected to modify the stability of uniform deposited layers, it will be studied on combining the kinetic equation (\ref{coveragedynCVD}) with (\ref{chempot}). The dynamics of the active species coverage then becomes:

\begin{equation}\label{CVDdynamics}
\partial_t c({\mathbf r},t) = R(c({\mathbf r},t))   -\vec\nabla \frac{D}{k_BT} \vec\nabla [  \epsilon_0c ({\mathbf r},t) - k_BT\ln \frac{c ({\mathbf r},t)}{ 1-c ({\mathbf r},t)} + \xi_0^2\nabla^2c ({\mathbf r},t)]
\end{equation}
where $ R(c({\mathbf r},t))=\alpha (1-c({\mathbf r},t))^nc({\mathbf r},t)^m - \beta c({\mathbf r},t)^r$.

Note that the case $n=1,m=0$ which corresponds to direct absorption, without precursor molecule dissociation, and in the absence of chemical reaction with the substrate, has already been extensively studied elsewhere \cite{walgraef_02b,marcel}.I will thus concentrate on more general cases, which are relevant for CVD.

\subsection{$m < r$}\label{mlessthanr}
Let me consider first the case $m < r$, where the reaction part of the dynamics admits a single stable uniform steady state $c_0$. The linear stability analysis of this steady state versus small nonuniform perturbations $\sigma ({\mathbf r},t)= c({\mathbf r},t)-c_0$ is studied through its linear evolution equation, with $D = D_0 c(1-c)$:

\begin{equation}\label{CVDlinstab}
 \partial_t \sigma ({\mathbf r},t) = -\Gamma \sigma ({\mathbf r},t)   -\vec\nabla \frac{D_0}{k_BT} \vec\nabla [  \epsilon_0 c_0(1-c_0) - k_BT + \xi_0^2c_0(1-c_0)\nabla^2 ]\sigma ({\mathbf r},t)
\end{equation}
or
\begin{equation}\label{CVDlinstab1}
 \partial_t \sigma ({\mathbf r},t) = -\Gamma \sigma ({\mathbf r},t) -  D_0[  (\mu - 1 )\nabla^2 \sigma ({\mathbf r},t) + \mu l^2\nabla^4\sigma ({\mathbf r},t)]
\end{equation}
where $\Gamma = - \frac{dR(c)}{dc}\vert_{c_0}$ , $\mu =  \frac{\hat T_c}{T}$, $ \hat T_c = 4T_cc_0(1-c_0)$, $l^2=\frac{\xi_0^2}{\epsilon_0}$. This gives, in Fourier transform:

\begin{equation}\label{CVDlinstab3}
\partial_t \sigma ({\mathbf { q}},t) = -[\Gamma  -  D_0 (  (\mu - 1 ) q^2 -  \mu l^2   q^4]\sigma ({\mathbf { q}},t)
\end{equation}
Instability may only occur if $T < \hat T_c (\mu >1)$ and its threshold is given by $ q_i^2 = \frac{\hat T_c  - T_i }{2l^2\hat T_c} $ and $\Gamma = \frac{D_0\hat T_c}{4l^2T_i} (\frac{\hat T_c  - T_i }{\hat T_c})^2$. Uniform steady states are thus unstable for

\begin{equation}\label{tempthreshB}
T < T_i = 4T_cc_0(1-c_0)[1 - 2\frac{\Gamma l^2}{ D_0} (  \sqrt{1 + \frac{ D_0}{ \Gamma l^2}}-1 )]=4T_cc_0(1-c_0)(1-\theta_i)
\end{equation}

\subsection{$m > r$}\label{mlargerthanr}

When $m > r$, the reaction part of the dynamics may admit two nontrivial uniform steady states and we will discuss how this can affect, even qualitatively pattern formation in the system. Let me consider, as an example, the case $n=1$, $m=2$, $r=1$, which correspond to $ R(c) = \alpha (1-c)c^2 - \beta c$. If $\frac{\beta}{\alpha}> \frac{1}{4}$, the only possible steady state is zero and no steady deposited layer forms, since evaporation dominates deposition. On the other hand, if $\frac{\beta}{\alpha}< \frac{1}{4}$ there are three possible uniform steady states, $c_0=0$ and $c_{\pm}= \frac{1}{2}(1\pm \sqrt{1-4\frac{\beta}{\alpha}})$.

The linear evolution of perturbations of the trivial steady state $c_0=0$ is given by

\begin{equation}\label{CVDdynamicslin2A}
\partial_tc({\mathbf r},t) =- \beta c({\mathbf r},t) + D_0\nabla^2 c({\mathbf r},t)
\end{equation}
showing that this state is linearly stable.

On the other hand, the linear evolution of perturbations of the nontrivial steady states $c_{\pm}$ is given by

\begin{eqnarray}\label{CVDdynamicslin2B}
\partial_t\sigma({\mathbf r},t) &=&+\alpha c_{\pm}(1- 2 c_{\pm})\sigma({\mathbf r},t) + D_0[1- \mu  - \mu l^2\nabla^2]\nabla^2  \sigma({\mathbf r},t)\nonumber  \\
\partial_t\sigma({\mathbf q},t) &=&\alpha c_{\pm}(1 - 2 c_{\pm} )\sigma({\mathbf q},t) + D_0q^2[\mu -1 - \mu l^2q^2  ]\sigma({\mathbf q},t)\nonumber\\
\end{eqnarray}
showing that $c_-$ is unstable since $ 1 - 2 c_{-}>0$, while $c_+$ is stable provided $\mu <1$. When $\mu >1$, it may become unstable versus nonuniform perturbations. For $c_+=\frac{1}{2}$, it is unstable for all perturbations with wavenumbers in the domain defined by $0<q^2<\frac{\mu -1}{\mu l^2}$, which extends to 0 as in spinodal decomposition. For $c_+>\frac{1}{2}$, this instability domain does not extend to zero. Since the maximum growth rate corresponds to $q^2=\frac{\mu -1}{2\mu l^2}$, the instability threshold is defined by $\alpha c_+(2 c_+ - 1 )= \frac{D_0}{4l^2\mu_i} (\mu_i-1)^2$, and $c_+$ is unstable for
\begin{equation}\label{tempthreshC}
T < T^+_i = 4T_cc_+(1-c_+)[ 1- \Lambda(  \sqrt{1 +\frac{2}{\Lambda}}- 1) ]= 4T_cc_+(1-c_+)(1-\theta^+_i)
\end{equation}
where $\Lambda=\frac{2l^2\alpha c_+(2 c_+ - 1 )}{D_0}$, and
\begin{equation}\label{unstableq}
\frac{1}{2}[\frac{(1-\mu)}{l^2}-\sqrt{(\frac{(1-\mu)}{l^2})^2 - \mu\Lambda}]< q^2 < \frac{1}{2}[\frac{(1-\mu)}{l^2}+\sqrt{(\frac{(1-\mu)}{l^2})^2 - \mu\Lambda}]
\end{equation}

Note that for diffusion induced by thermally activated atomic jumps, the diffusion coefficient behaves as $D_0 = D^*_0\exp -(\Omega /T)$, and its temperature dependence may be discarded for $T>>\Omega$. Since physical parameters of the model are highly sensitive to interatomic potentials, lattice coordination number, etc. , the validity of this approximation has to be considered carefully. Furthermore, for $T<\Omega$, diffusion coefficients strongly decrease with decreasing temperature, so that relaxation finally dominates the dynamics, which rules out any instability at sufficiently low temperatures. Effectively, at low temperatures, diffusion strongly decreases, and, for $D_0 \simeq D^*_0\exp -(\Omega /T)$ ($T<\Omega$), the instability condition (\ref{tempthreshB}) may be expanded in powers of $D^*$, which yields (cf \cite{walgraef_02b}):

\begin{equation}\label{lowTinstability}
\frac{(\mu -1)^2}{\mu} \exp -(\Omega^* \mu)> \frac{4l^2\alpha c_+(2 c_+ - 1 )}{D^*_0}
\end{equation}
with $\Omega^*=\frac{\Omega}{4T_cc_+(1-c_+)}$. For small $\Omega^*$, this condition is satisfied, for $\Omega^*<\frac{D^*_0}{4el^2\alpha c_+(2 c_+ - 1 )}$, in a finite temperature range, with an upper and lower bound, which thus excludes pattern forming instability at sufficiently low temperatures. The weakly nonlinear analysis, performed in the next section to determine the dynamically selected patterns, is thus appropriate for systems with high atomic mobility ($\Omega^* << 1$), and temperatures slightly below $T_i$.

\section{Weakly nonlinear dynamics and pattern selection beyond instability.}\label{nonlinear}

The evolution equation for perturbations $\sigma ({\mathbf r},t)$ of a uniform steady state $c_0$, derived from eq. (\ref{CVDdynamics}), may be written as
\begin{equation}\label{CVDnolin}
 \partial_t \sigma ({\mathbf r},t) = -{\cal R}(c_0, \sigma ({\mathbf r},t))   -\vec\nabla \frac{D(c({\mathbf r},t))}{k_BT} [  \epsilon_0   - \frac{k_BT }{c ({\mathbf r},t)(1-c ({\mathbf r},t))} + \xi_0^2\nabla^2 ]\vec\nabla \sigma ({\mathbf r},t)
\end{equation}
Close to instability, it may be expanded in a series expansion in powers of $\sigma ({\mathbf r},t)$, limited to the first nonlinear stabilizing term, which usually corresponds to cubic nonlinearities. This weakly nonlinear dynamics is a reduction of the full dynamics which leads to a dramatic simplification of it, that nevertheless captures the asymptotic properties of the system's evolution and allows an accurate description of pattern formation, selection and stability up to a finite distance beyond instability \cite{ghoniemwalgraef1}. Such weakly nonlinear analysis will now be performed for explicit expressions of the diffusion coefficient and different types of reaction processes.

\subsection{Weakly nonlinear analysis for $D = D_0 c(1-c)$ and $m<r$}\label{wnlmlessthanr}
In this case, the weakly nonlinear evolution equation, derived from eq. (\ref{CVDnolin}), and limited to cubic nonlinearities writes:
\begin{eqnarray}\label{CVDnolin2}
\partial_t \sigma ({\mathbf r},t) &=& - \Gamma \sigma ({\mathbf r},t) -  D_0 [ (\mu - 1 )\nabla^2 \sigma ({\mathbf r},t)+  \mu l^2\nabla^4\sigma ({\mathbf r},t)]\nonumber\\
&-& \Gamma_2 \sigma^2 ({\mathbf r},t)-  D_2\mu \vec{\nabla }[\sigma ({\mathbf r},t) . \vec{\nabla} (\sigma ({\mathbf r},t)+ l^2\nabla^2\sigma ({\mathbf r},t))]\nonumber\\
&-& \Gamma_3 \sigma^3 ({\mathbf r},t) + D_3\mu\vec{\nabla }[  \sigma^2 ({\mathbf r},t).\vec{\nabla}( \sigma ({\mathbf r},t)+ l^2\nabla^2\sigma ({\mathbf r},t))]
\end{eqnarray}
where $\Gamma = - \frac{dR(c)}{dc}\vert_{c_0}$, $\Gamma_2 = - \frac{1}{2}\frac{d^2R(c)}{dc^2}\vert_{c_0}$, $\Gamma_3 = - \frac{1}{6}\frac{d^3R(c)}{dc^3}\vert_{c_0}$, $ D_2 = \frac{ D_0(1-2c_0)}{ c_0(1-c_0)}$, $D_3 = \frac{D_0}{ c_0(1-c_0)}$. In this notation, $\mu$ will be chosen as the bifurcation parameter. It is the only parameter  which explicitly depends on temperature, while the different diffusion coefficients depend on the steady state coverage.

As discussed in the preceding section, uniform coverage is unstable for $\mu > \mu_i$ versus nonuniform perturbations. Fourier modes with $q=q_i$ have the maximum growth rate and they may be expected to dominate the system's evolution and to generate spatial patterns. Close to instability, pattern evolution and selection may be studied through their amplitude equations \cite{walgraef_96,ghoniemwalgraef1}. In this analysis, the simplest autocatalytic processes will be considered, where ($n=1$, $m=1$, $r=2$, $c_0=\frac{\alpha}{\alpha + \beta}$, $\Gamma =\alpha $, $\Gamma_2=\alpha  + \beta  $, $\Gamma_3=0$, and $n=1$, $m=1$, $r=1$, with $c_0=\frac{\alpha - \beta}{\alpha  }$, $\Gamma =\alpha - \beta$, $\Gamma_2=\alpha  $, $\Gamma_3=0$ ) to allow an explicit analysis of the balance between reaction and diffusion generated nonlinearities. The different types of patterns which may arise from this dynamics, and their stability, will now be discussed.

\subsubsection{Stripes}
The simplest patterns, able to develop on an isotropic substrate, correspond to stripes, such that $\sigma ({\mathbf r},t)= A_1({\mathbf r},t)e^{iq_ix} + A^*_1({\mathbf r},t)e^{-iq_ix}$, where $q_1=q_i$. On neglecting higher harmonics, their amplitude equation derived from eq. (\ref{CVDnolin2}) writes:

\begin{equation}
\tau\partial_t A_1 ({\mathbf r},t) = \epsilon A_1({\mathbf r},t)
+\lambda(  \nabla_x + \frac{ i}{2q_i} \nabla_y^2  )^2A_1({\mathbf r},t)
-  \kappa\vert A_1 ({\mathbf r},t)\vert^2 A_1 ({\mathbf r},t))
\end{equation}
where $\epsilon = \frac{\mu - \mu_i}{\mu_i}=\frac{ T_i - T}{T}$, $ \tau=\frac{4l^2\mu_i}{D_0(\mu_i^2-1)}=\frac{(\mu_i-1)}{\alpha (1+\mu_i)}$ (or $\frac{(\mu_i-1)}{(\alpha -\beta)(1+\mu_i)}$ for $r=1$), $\lambda =\frac{8l^2\mu_i}{\mu_i-1}$ and $\kappa = \frac{\mu}{\mu_ic_0(1-c_0) }$. As in \cite{walgraef_02b}, steady state stripe patterns of amplitude $\vert A_1 \vert^2 = A_0^2 = \frac{ T_i - T}{ T_i}c_0(1-c_0)$ should develop for any $T<T_i$. However, these patterns are unstable versus hexagonal ones in a finite domain of this temperature range as discussed in \cite{ghoniemwalgraef1} and in the next section.

\subsubsection{Hexagons }

Effectively, hexagonal planforms, such that
\begin{equation}
\sigma ({\mathbf r},t)= A_1({\mathbf r},t)e^{i\vec q_1\vert r} + A_2({\mathbf r},t)e^{i\vec q_2\vert r} + A_3({\mathbf r},t)e^{i\vec q_3\vert r} + c.c.,
 \end{equation}
 where $\vec q_1 +\vec q_2 + \vec q_3 =0$ and $\vert\vec q_1\vert = \vert\vec q_2\vert = \vert\vec q_3\vert = q_i$, are described by the following amplitude equations:

\begin{eqnarray}\label{hexamplitude}
\tau\partial_t A_1 &=& \epsilon A_1 + \frac{\lambda}{q_i^2}(\vec q_1.\vec\nabla)^2A_1 + 2\nu A^*_2A^*_3-  \kappa A_1(\vert A_1\vert^2 +\rho\vert  A_2 \vert^2+ \rho\vert  A_3 \vert^2)\nonumber \\
\tau\partial_t A_2 &=& \epsilon A_2 + \frac{\lambda}{q_i^2}(\vec q_1.\vec\nabla)^2A_2 + 2\nu A^*_3A^*_1-  \kappa A_2(\vert A_2\vert^2 +\rho\vert  A_1 \vert^2+ \rho\vert  A_3 \vert^2)\nonumber \\
\tau\partial_t A_3 &=& \epsilon A_3 + \frac{\lambda}{q_i^2}(\vec q_1.\vec\nabla)^2A_3 + 2\nu A^*_2A^*_1-  \kappa A_3(\vert A_3\vert^2 +\rho\vert  A_2 \vert^2+ \rho\vert  A_1 \vert^2)\nonumber \\
\end{eqnarray}
where $\nu = - \tau(\alpha  + \beta  ) + \frac{\mu(1-2c_0)}{ \mu_ic_0(1-c_0)}$ and $\rho=2$ (or $\nu = - \tau\alpha  + \frac{\mu(1-2c_0)}{ \mu_ic_0(1-c_0)}$ for $r=1$). Common analysis shows that these equations admit steady state solutions corresponding to hexagonal planforms of amplitude
\begin{equation}
\vert A_1\vert = \vert A_2\vert = \vert A_3\vert = R  = \frac{1}{5\kappa}\biggl(\nu + \sqrt{\nu^2 + 5\kappa\epsilon }\biggr)
 \end{equation}
 Hence, these solutions may appear subcritically for $\epsilon > - \frac{\nu^2 }{5\kappa}$ and are stable provided $\epsilon < \frac{16\nu^2 }{\kappa}$   \cite{ghoniemwalgraef1}. Their existence and stability range is thus given by
 \begin{equation}
T_{h-}= T_i - 16\Delta  T_i < T  <T_i +\frac{\Delta  T_i}{5}=T_{h+}
  \end{equation}
  where $\Delta  T_i = \frac{T_i\nu^2 }{\kappa}\simeq 4T_c(1-\theta_i)(  \frac{ \alpha-\beta}{\alpha  + \beta })^2(\mu_i\simeq 1)$. The sign of $\nu$ is important since it determines the type of hexagons which are stable in this domain. Effectively, for $\nu >0$ they correspond to $H_+$ hexagons (with the amplitude maxima at the hexagon centers), while for $\nu <0$, they correspond to $H_-$ hexagons (with the amplitude minima at the hexagon centers)\cite{ghoniemwalgraef1}. Since in this example, $1-2c_0 = \frac{\beta -  \alpha}{\alpha  + \beta } $, $\nu$ is negative for $\beta    < \alpha$, i.e. for high coverage, which results in a pattern formed by an equilateral triangular lattice of low coverage dots ($H_-$ hexagons). If $(1-2c_0)>0$, or $\beta    > \alpha$, i.e. for low coverage, $\nu$ is positive only if $\frac{\alpha}{\beta} < 1 - \frac{\mu_i(\mu_i-1)}{\mu (\mu_i+1)}$ (or if $\frac{\alpha}{\beta} < 2 - \frac{\mu_i(\mu_i-1)}{\mu (\mu_i+1)}$for $r=1$). The limit between $0$ and $\pi$ hexagons is thus shifted by reactive terms of the dynamics which favor the formation of $H_-$ hexagons.

As mentioned in the preceding section, striped patterns may be unstable versus hexagonal planforms. Effectively, the linear evolution of hexagonal perturbations of stripes such that $\sigma ({\mathbf r},t)= A_0({\mathbf r},t)e^{iq_ix} + A^*_0({\mathbf r},t)e^{-iq_ix}$ is given by

\begin{eqnarray}
\tau\partial_t A_2 &=& \epsilon A_2 + 2\nu A^*_3A^*_0-  2\kappa A_2\vert  A_0 \vert^2= - \epsilon A_2 + 2\nu A^*_3A^*_0\nonumber \\
\tau\partial_t A^*_3 &=& \epsilon A^*_3 + 2\nu A_2A_0-  2\kappa A^*_3\vert  A_0 \vert^2=-\epsilon A^*_3 + 2\nu A_2A_0\nonumber \\
\end{eqnarray}
and the eigenvalues of the corresponding evolution matrix are given by $\omega = - \epsilon \pm 2\nu\sqrt{\frac{\epsilon}{\kappa}}$ and stripes are unstable  $0 < \epsilon < \frac{4\nu^2}{\kappa}$.

As a result, hexagons and stripes are simultaneously stable in the bistability domain defined by
$ \frac{4\nu^2}{\kappa}< \epsilon < \frac{16\nu^2 }{\kappa}$ or $T_{h-}= T_i - 16\Delta  T_i < T  <T_i - 4\Delta  T_i=T_{s}$.

\subsubsection{Squares}

For squares patterns, such that $\sigma ({\mathbf r},t)= A_1({\mathbf r},t)e^{iq_ix}+ B_1({\mathbf r},t)e^{iq_iy}+ c.c.$, the corresponding amplitude equations are:

\begin{eqnarray} \tau\partial_t A_1 &=& \epsilon + \lambda(  \nabla_x + \frac{ i}{2q_i} \nabla_y^2  )^2A_1 -  \kappa A_1(\vert A_1\vert^2 +\rho\vert  B_1 \vert^2)\nonumber \\ \tau\partial_t B_1 &=& \epsilon B_1 + \lambda(  \nabla_y + \frac{ i}{2q_i} \nabla_x^2  )^2B_1- \kappa B_1(\vert B_1\vert^2+\rho\vert A_1\vert^2)\end{eqnarray}
Since $\rho=2 $, square patterns are always unstable when described by such equations.

Pattern formation phenomena, discussed in this section, do not qualitatively differ from the case described in \cite{walgraef_02b}. The effect of more complex reaction processes associated to chemical vapor deposition is only quantitative and modifies the existence and stability ranges of the different patterns, according to the relative importance of reaction and diffusion-induced nonlinearities.

\subsection{Weakly nonlinear analysis for $D = D_0 c(1-c)$ and $m>r$.}\label{mgreater}

Let us now consider the case where deposition processes are autocatalytic, and with a higher degree of nonlinearity than desorption ones.  Corresponding uniform steady states and their linear stability have been described previously. As far as the nonlinear dynamics is concerned, two cases have to be considered.

\subsubsection{The case of critical coverage ($c_+  =\frac{1}{2}$ with $n=1$, $m=2$, $r=1$).}
For critical coverage, $2 c_+ - 1 =0$, and the nonlinear evolution of coverage perturbations writes:

\begin{eqnarray}\label{CVDdynamics2crit}
\partial_t\sigma({\mathbf r},t) &=& - D_0[\mu - 1  +\mu l^2\nabla^2 ]\nabla^2 \sigma ({\mathbf r},t)
-\frac{\alpha}{2}\sigma^2({\mathbf r},t)  \nonumber  \\
&-&\alpha  \sigma^3({\mathbf r},t) +4D_0\mu[  1   + l^2\nabla^2 ]\vec\nabla \sigma^2({\mathbf r},t)\vec\nabla\sigma ({\mathbf r},t)
\end{eqnarray}
In Fourier transform, for $\mu >1$, the band of unstable wavenumbers, $q$, is defined by $ 0 < q^2 < \frac{\mu - 1 }{\mu l^2}$, which includes zero as in spinodal decomposition. Hence, the spatial gradient terms of the dynamics should induce coarsening of developing spatial patterns. However the remain nonlinear terms should induce refining of the structure. The balance between coarsening and refining should, at the end, stabilize nanostructures with finite wavenumber, corresponding to stripes or hexagons due to the presence of quadratic nonlinearities in the dynamics. Nevertheless, due to the presence of a nonlinearly damped zero mode in the dynamics, pattern and wavelength selection should be studied numerically.

\subsubsection{The case of high coverage ($ c_+ \lesssim 1$ with $n=1$, $m=2$, $r=1$). }

For high coverage, $2 c_+ - 1 \ne 0$, the nonlinear evolution of coverage perturbations writes:
\begin{eqnarray}\label{CVDdynamics2B}
\partial_t\sigma({\mathbf r},t) &=&+\alpha c_+(1- 2 c_+)\sigma  -D_0[\mu - 1 + \mu l^2\nabla^2 ]\nabla^2 \sigma ({\mathbf r},t)\nonumber\\
&+&\alpha(1 - 3c_+)\sigma^2  - \frac{D_0\mu (1-2c_+)}{c_+(1-c_+)} [   1  + l^2\nabla^2 ]\vec\nabla[\sigma({\mathbf r},t))\vec\nabla\sigma ({\mathbf r},t)  \nonumber  \\
&-&\alpha  \sigma^3 +\frac{D_0\mu}{ c_+(1-c_+)}[  1  + l^2\nabla^2 ]\vec\nabla \sigma^2({\mathbf r},t))\vec\nabla\sigma ({\mathbf r},t)
\end{eqnarray}
and in Fourier transform, the evolution of a perturbation of critical wavevector is given by

\begin{eqnarray}\label{CVDnolin2B}
&&\partial_t\sigma({\mathbf q_i},t) =\frac{D_0(\mu_i^2-1)}{4l^2\mu_i}\frac{\mu - \mu_i}{\mu_i} \sigma({\mathbf q_i},t)\nonumber\\
&+&\int d{\mathbf k}[\alpha(1 - 3c_+ ) +   \frac{D_0\mu ( 1 + \mu_i)}{2\mu_ic_+(1-c_+)} (1-2c_+)({\mathbf q_i}.{\mathbf k})]\sigma({\mathbf {q_i-k}},t)\sigma({\mathbf k},t)\nonumber  \\
&-&\int d{\mathbf k}\int d{\mathbf k_1}[\alpha   +\frac{D_0\mu ( 1 + \mu_i)}{2\mu_i c_+(1-c_+)} ]({\mathbf q_i}.{\mathbf k_1}) ]\sigma({\mathbf {q_i-k}},t)\sigma({\mathbf {k-k_1}},t)\sigma({\mathbf k_1},t)\nonumber  \\
\end{eqnarray}
Amplitude equations for different types of spatial patterns may then be derived from this equation, allowing to assess their existence and stability, as in the discussion performed in section \ref{wnlmlessthanr}.

\subsubsection{Stripes}

Here also, stripes will be considered first as the simplest structure to arise from this dynamics. Their amplitude equations may be written, at lowest order in $\epsilon = \frac{\mu - \mu_i}{\mu_i}$. as:
\begin{equation}\label{CVDstripesB}
\tau_+\partial_t A_1 ({\mathbf r},t) = [\epsilon A_1
+\lambda_+(  \nabla_x + \frac{ i}{2q_i} \nabla_y^2  )^2]A_1({\mathbf r},t)
- u_+(q_i^2)\vert A_1 ({\mathbf r},t)\vert^2 A_1 ({\mathbf r},t))
\end{equation}
where
$\tau_+ = \frac{4l^2\mu_i}{D_0(\mu_i^2-1)}$, $\alpha_+= \frac{1}{c_+(2 c_+ - 1 )} \frac{(\mu_i-1)}{(1+\mu_i)}$ and $\kappa_+({\mathbf q_i}.{\mathbf k_1})=  \frac{\mu}{\mu_ic_+(1-c_+)}\frac{ ({\mathbf q_i}.{\mathbf k_1})}{q_i^2}$ and $u_+(q_i^2)=3\alpha_+ + \kappa_+(q_i^2)$. Steady stripes of amplitude given by $\vert A_0\vert^2= \frac{\epsilon}{3 \alpha_+ + \kappa_+}$ may thus appear supercritically for $\epsilon >0$.

\subsubsection{Hexagons}
On the other hand, amplitude equations for hexagonal or honeycomb planforms write:
\begin{eqnarray}\label{hexamplitude2}
\tau_+\partial_t A_1 &=& [\epsilon + \frac{\lambda_+}{q_i^2}(\vec q_1.\vec\nabla)^2]A_1 + 2\nu_+ A^*_2A^*_3-  u_+ A_1[\vert A_1\vert^2 +\rho(\vert  A_2 \vert^2+ \vert  A_3 \vert^2)]\nonumber \\
\tau_+\partial_t A_2 &=& [\epsilon  + \frac{\lambda_+}{q_i^2}(\vec q_1.\vec\nabla)^2]A_2 + 2\nu_+ A^*_3A^*_1-  u_+ A_2[\vert A_2\vert^2 +\rho(\vert  A_1 \vert^2+ \vert  A_3 \vert^2)]\nonumber \\
\tau_+\partial_t A_3 &=& [\epsilon  + \frac{\lambda_+}{q_i^2}(\vec q_1.\vec\nabla)^2]A_3 + 2\nu_+ A^*_2A^*_1-  u_+ A_3[\vert A_3\vert^2 +\rho(\vert  A_2 \vert^2+\vert  A_1 \vert^2)]\nonumber \\
\end{eqnarray}
where $\nu_+$ and $u_+$ are taken at $q_i^2$, and $\nu_+(q_i^2)= \frac{(1 - 3c_+ )}{ c_+(2 c_+ - 1)}\frac{(\mu_i-1)}{(1+\mu_i)} +  \frac{\mu(1-2c_+)}{\mu_ic_+(1-c_+)}$,  and $\rho=2$. It turns out that $\nu_+(q_i^2)$ is always negative, which leads to the formation of $\pi$ or ($-$) hexagons. Furthermore, according to standard pattern formation theory, these hexagons are stable in the range

\begin{equation}
T^+_{h-}= T^+_i - 16\Delta  T^+_i < T  <T^+_i +\frac{\Delta  T^+_i}{5}=T^+_{h+}
  \end{equation}
  where $\Delta  T^+_i = \frac{T^+_i\nu_+^2 }{3 \alpha_+ +\kappa_+}$, while stripes are stable in the range $  T  <T^+_i - 4\Delta  T^+_i=T^+_{s}$. As a result, hexagons and stripes are simultaneously stable in the bistability domain defined by $ T^+_{h-}=T^+_i - 16\Delta  T^+_i < T  <T^+_i - 4\Delta  T^+_i=T_{s}$.

\subsubsection{Pattern Selection}

Pattern selection resulting from the results obtained so far in section \ref{mgreater} may then be summarized as follows.

\begin{figure}
\includegraphics[width=4.5in]{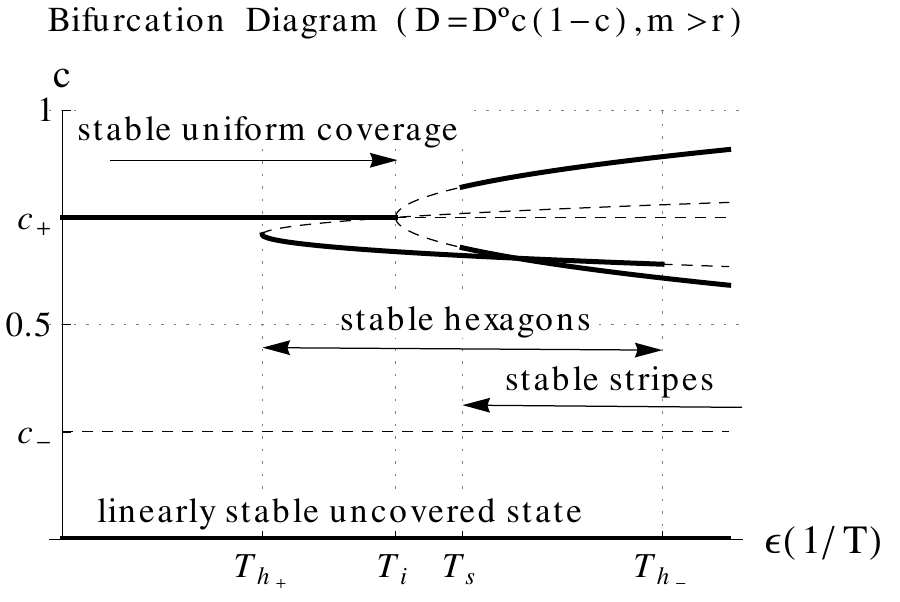}
 \caption{Schematic bifurcation diagram associated with eq. (\ref{CVDnolin2B}) in the coverage-temperature plane. Heavy lines represent stable steady states and dotted lines represent unstable ones. According to the temperature one may observe different multistability domains which should allow the existence of various types of localized structures.} \label{figure1}
\end{figure}

\begin{enumerate}
\item if $\frac{\beta}{\alpha}> \frac{1}{4}$ the only possible steady state is zero and no deposited layer forms since evaporation dominates deposition.

\item if $\frac{\beta}{\alpha}< \frac{1}{4}$ there are three possible uniform steady states, $c_0=0$ and $c_{\pm}= \frac{1\pm \sqrt{1-4\frac{\beta}{\alpha}}}{2}$.
\begin{itemize}
\item for $T>T_i$, $c_0$ and $c_+$ are stable while $c_-$ is unstable. In this bistability regime, localized structures corresponding to covered islands or dots in an uncovered background are possible.

\item for $T<T_i$, $c_0$ remain stable and $c_-$ unstable. $c_+$ becomes unstable versus spatial patterns corresponding to stripes or hexagons. $\pi$  hexagons may appear subcritically for $T  <T^+_{h+}$ and are stable for $ T^+_{h-} < T  <T^+_{h+}$. Stripes, on the other hand are stable for $  T  <T_{s}$. Hence hexagons, stripes and the uncovered state may be simultaneously stable, which may lead to several types of coexisting or localized patterns, such as islands or dots of patterned states in the uncovered background. These results are illustrated in the schematic bifurcation diagram presented in figure \ref{figure1}.
\end{itemize}
\end{enumerate}
We have thus found that reactive autocatalytic terms in precursor decomposition favor the formation of $H_-$ hexagonal patterns in the growing layer (similar to nanomesh structures) at low temperatures, which is reminiscent of a Frank-Van der Merwe growth with nanostructures. At high temperatures, however, localized island formation, which should initiate a Volmer-Weber type of film growth, can occur.

\subsection{Weakly nonlinear analysis with $D_0 = D^*_0 \exp -(\Omega /T)$ and $m>r$..}\label{Dtempdep}

When atomic transport in the deposited layer is due thermally activated atomic jumps, their activation energy has to be taken into account, and the diffusion coefficient may be written as $D = D^*_0 c(1-c)\exp -(\Omega /T)$, where $\Omega$ is proportional to this activation energy. Equation (\ref{CVDdynamics2B}) then becomes:

\begin{eqnarray}\label{CVDdynamics2C}
&&\partial_t\sigma({\mathbf r},t) =+\alpha c_+(1- 2 c_+)\sigma  -D^*_0\exp -(\Omega^* \mu)[\mu - 1 + \mu l^2\nabla^2 ]\nabla^2 \sigma ({\mathbf r},t)\nonumber\\
&+&\alpha(1 - 3c_+)\sigma^2  - \frac{D^*_0\exp -(\Omega^* \mu)\mu (1-2c_+)}{c_+(1-c_+)} [   1  + l^2\nabla^2 ]\vec\nabla[\sigma({\mathbf r},t))\vec\nabla\sigma ({\mathbf r},t)  \nonumber  \\
&-&\alpha  \sigma^3 +\frac{D^*_0\exp -(\Omega^* \mu)\mu}{ c_+(1-c_+)}[  1  + l^2\nabla^2 ]\vec\nabla \sigma^2({\mathbf r},t))\vec\nabla\sigma ({\mathbf r},t)
\end{eqnarray}
Since the maximum of the linear growth rate corresponds to wavevectors such that $q^2= q_m^2 = \frac{\mu - 1 }{2\mu l^2}$, the evolution equation for such modes writes, in Fourier transform:

\begin{eqnarray}\label{CVDnolin2C}
&&\partial_t\sigma({\mathbf q_m},t) =+ \,\, [\alpha c_+(1- 2 c_+)  +D^*\frac{(\mu - 1)^2 }{\mu} ] \sigma ({\mathbf q_m},t)\nonumber\\
&+&\int d{\mathbf k}[\alpha(1 - 3c_+ ) +   \frac{D^*(\mu^2 -1) (1-2c_+)}{\mu c_+(1-c_+)} \frac{({\mathbf q_m}.{\mathbf k})}{q_m^2}]\sigma({\mathbf {q_m-k}},t)\sigma({\mathbf k},t)\nonumber  \\
&-&\int d{\mathbf k}\int d{\mathbf k_1}[\alpha   +\frac{D^*(\mu^2 -1)}{\mu c_+(1-c_+)}\frac{({\mathbf q_m}.{\mathbf k_1})}{q_m^2} ]\sigma({\mathbf {q_m-k}},t)\sigma({\mathbf {k-k_1}},t)\sigma({\mathbf k_1},t)\nonumber  \\
\end{eqnarray}
where $D^* = \frac{D^*_0}{4l^2}e^{-(\Omega^* \mu)}$. As a result, instability occurs for
\begin{equation}\label{lowTinstability2}
\Sigma_+=\frac{(\mu -1)^2}{\mu} \exp -(\Omega^* \mu)> \frac{4l^2\alpha c_+(2 c_+ - 1 )}{D^*_0}=\Sigma_-
\end{equation}
This condition is satisfied, for
\begin{equation}\frac{4l^2\alpha c_+(2 c_+ - 1 )}{D^*_0}< Max (\frac{(\mu -1)^2}{\mu} e^{-\Omega^* \mu})= \frac{(\mu^* -1)^2}{\mu^*} e^{-\Omega^* \mu^*}\simeq \frac{1}{\Omega^*e}
\end{equation}
for small $\Omega^* $  since, in this limit, $\mu^* = \frac{1+\Omega^*}{2\Omega^*}[1+\sqrt{1+ \frac{4\Omega^*}{(1+\Omega^*)^2}}]$
 (the case of larger $\Omega^* $ will be discussed later). As mentioned earlier, when this condition is satisfied, uniform layers are unstable in a finite temperature range, with an upper ($T_{i+}$) and lower ($T_{i-}$) bound. This can be illustrated explicitely for $ \frac{4l^2\alpha c_+(2 c_+ - 1 )\Omega^*e}{D^*_0}\lesssim 1 $. In this case, the linear growth rate of (\ref{CVDnolin2C}) may be approximated by

\begin{eqnarray}\label{lingrowth2}
\alpha c_+(1- 2 c_+)  &+&D^*\frac{(\mu^* - 1)^2 }{\mu^*} - D^*\frac{(\mu^*)^2 + 2\mu^*-1}{2(\mu^*)^3}e^{-(\Omega^* \mu^*)}(\mu - \mu^*)^2\nonumber\\
&=&-\Phi [(\frac{\mu - \mu^*}{\mu^*})^2 -\delta^2]
\end{eqnarray}
where $\Phi^* = \frac{D^*_0}{4l^2}e^{-(\Omega^* \mu^*)}\frac{(\mu^* - 1)^2 }{\mu^*}\simeq \frac{D^*_0}{4e\Omega^* l^2}$, $\delta ^2 = 1 - \frac{4l^2\alpha c_+(2 c_+-1) \mu^*}{D^*_0 (\mu^* - 1)^2 }\,\,e^{(\Omega^* \mu^*)}\simeq 1- \frac{4e\Omega^*l^2\alpha c_+(2 c_+-1) }{D^*_0  }$ and instability occurs for $\mu_-=\mu^*(1 -\delta) < \mu < \mu^*(1 +\delta )=\mu_+$ or $T_{i-}=\frac{T^*}{1 +\delta}<T<\frac{T^*}{1 -\delta}=T_{i+}$.

Eq. (\ref{CVDnolin2C}) may then be rewritten as:

\begin{eqnarray}\label{CVDnolin2D}
&&\tau^*\partial_t\sigma({\mathbf q_m},t) = \,\,  \frac{(\mu_+ - \mu)(\mu - \mu_-)}{(\mu^*)^2} \sigma ({\mathbf q_m},t)\nonumber\\
&+&\int d{\mathbf k}[\nu^*(q_m,k)]\sigma({\mathbf {q_m-k}},t)\sigma({\mathbf k},t)\nonumber  \\
&-&\int d{\mathbf k}\int d{\mathbf k_1}[\kappa^* (q_m,k_1)]\sigma({\mathbf {q_m-k}},t)\sigma({\mathbf {k-k_1}},t)\sigma({\mathbf k_1},t)\nonumber  \\
\end{eqnarray}
where $\tau^* = \frac{1 }{\Phi^*  }$, $\nu^*(q_m,k)= \frac{\alpha(1-3c_+)}{\Phi^* } +   \frac{({\mu^*}^2 -1) (1-2c_+)}{(\mu^* - 1)^2 c_+(1-c_+)} \frac{({\mathbf q_m}.{\mathbf k})}{q_m^2}$ and $ \kappa^* (q_m,k_1)= \frac{\alpha  }{\Phi^*  } +   \frac{({\mu^*}^2 -1)}{(\mu^* - 1)^2 c_+(1-c_+)}\frac{({\mathbf q_m}.{\mathbf k_1})}{q_m^2}$.
\subsubsection{Stripes}
In this case, amplitude equation analysis shows that, for each $\mu$ in the range $\mu_-<\mu<\mu_+$, stripes of amplitude
$\vert A_m\vert^2= \frac{(\mu_+ - \mu)(\mu - \mu_-)}{(\mu^*)^2}\frac{(2c_+-1)(1-c_+)}{2 - c_+}$ and wavenumber $q_m^2 = \frac{\mu - 1 }{2\mu l^2}$ are the preferred one-dimensional patterns since the coefficient of the cubic coupling term of the stripes amplitude equation is given by $\frac{12e\Omega^* l^2\alpha}{D_0^* }+ \frac{1}{c_+(1-c_+)}\simeq \frac{2 - c_+}{(2c_+-1)c_+(1-c_+)}$, for small $\Omega^*$.

\subsubsection{Hexagons}
\begin{figure}
\includegraphics[width=4.5in]{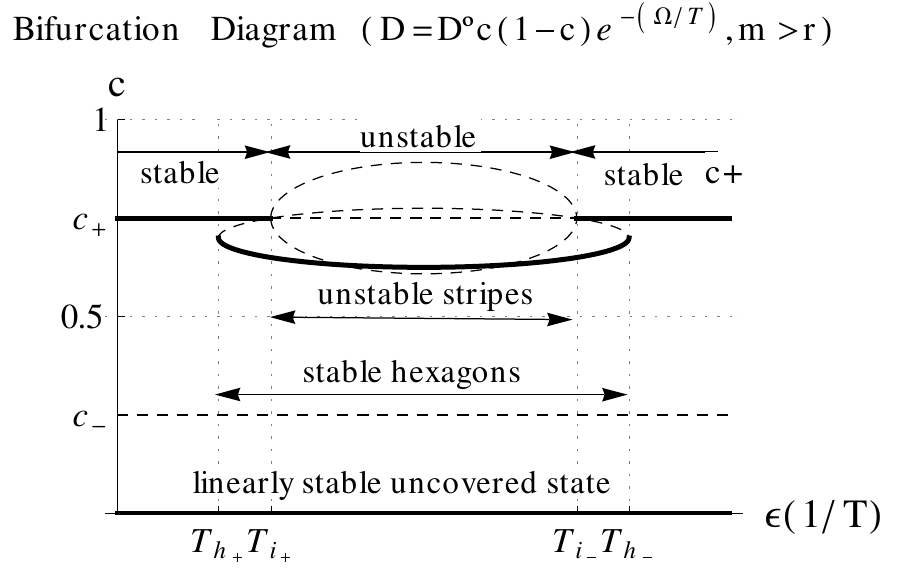}
 \caption{Schematic bifurcation diagram associated with eq. (\ref{CVDnolin2C}) in the coverage-temperature plane for $\delta^2 < \frac{4c_+(1-c_+)}{(2 - c_+)(2c_+-1)}$. Heavy lines represent stable steady states and dotted lines represent unstable ones. According to temperature, uniform deposited layer solution, self-assembled hexagonal patterns and uncovered substrate may be simultaneously stable. } \label{figure2}
\end{figure}

In the same conditions, the coefficient of quadratic nonlinearities in the amplitude equations for hexagonal patterns  is given by $ \nu^*(q_m^2)\simeq \frac{4e\Omega^* l^2\alpha(1-3c_+ )}{D_0^* }+ \frac{1-2c_+}{c_+(1-c_+)}\simeq \frac{1}{1-2c_+}$. Since it is always negative, a similar analysis than the one performed in previous sections leads to the fact that negative or $\pi$ hexagons exist and appear subcritically for

\begin{equation}
(\frac{\mu - \mu^*}{\mu^*})^2 < \delta^2 + \frac{c_+(1-c_+)}{5(2 - c_+)(2c_+-1)}= \delta_{h}^2
\end{equation}
or
\begin{equation}
\mu^*(1 -\delta_{h}) < \mu < \mu^*(1 +\delta_{h} )
\end{equation}
or
\begin{equation}
T_{h-}=\frac{T^*}{1 +\delta_{h}} < T < \frac{T^*}{1 -\delta_{h}}=T_{h+}
\end{equation}
On the other hand, stripes are only stable for
\begin{equation}
(\frac{\mu - \mu^*}{\mu^*})^2 <   \delta^2 - \frac{4c_+(1-c_+)}{(2 - c_+)(2c_+-1)}= \delta_{s}^2
\end{equation}
or
\begin{equation}
\mu^*(1 -\delta_{s}) < \mu < \mu^*(1 +\delta_{s} )
\end{equation}or
\begin{equation}
T_{s-}=\frac{T^*}{1 +\delta_{s}} < T < \frac{T^*}{1 -\delta_{s}}=T_{s+}
\end{equation}
It turns out that if $\delta^2 < \frac{4c_+(1-c_+)}{(2 - c_+)(2c_+-1)}$, i.e. sufficiently close to threshold, stripes are always unstable and hexagons are stable. As a result, for systems with high atomic mobility ($\Omega^* << 1$), pattern forming instability may occur in a finite temperature range and should be excluded at sufficiently low temperatures as illustrated in the schematic bifurcation diagram presented in figure \ref{figure2}. In this temperature range, bistability should occur between hexagonally structured deposited layers and uncovered substrate. This could allow the formation of deposited localized islands of hexagonal self-assembled structures on the substrate.

\section{Discussion.}\label{discussion}

Having proposed an instability mechanism for coverage pattern formation in monolayers adsorbed on a substrate via chemical vapor deposition, it is of course essential to check whether this mechanism is experimentally possible. The case of $Al$, $Cu$, $Ti$ or $TiN$ films deposited via adsorption-desorption mechanisms only have been discussed in \cite{walgraef_02b}. However, due to the evergrowing number of experimental processes developed in this field and the impossibility to consider every single experimental system, I will just illustrate the results obtained so far with realistic values of experimental parameters. This may then trigger interest to perform more focused analysis.  To do so, let me consider values in the range of the ones which are commonly used in thin-film deposition.

\begin{figure}
\includegraphics[width=4.5in]{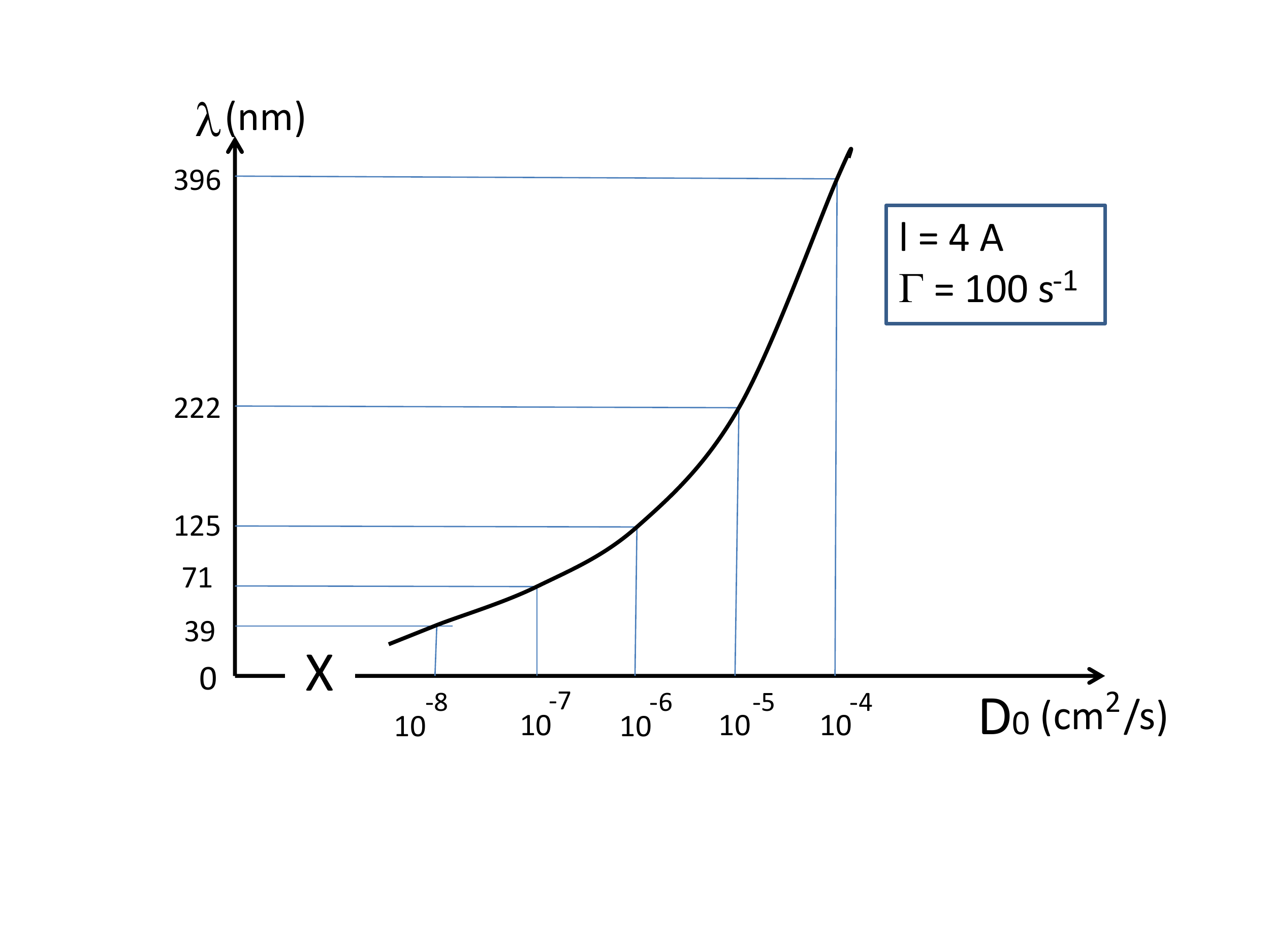}
 \caption{Representation of the critical wavelength dependence on the diffusion coefficient ($\lambda_i \simeq (\frac{l^2D_0}{\Gamma})^{1/4} $), for $l=4$ \AA\, $\Gamma = 10^{2}  s^{-1} $, when $D=D_0c(1-c)$ .} \label{figure3}
\end{figure}

For example, for a species crystallizing in FCC lattices, with a lattice constant $l=5$ \AA\, a pair interaction potential $\epsilon \simeq  0.15$ eV and deposited on (100) surfaces, with a very small lattice mismatch, $\epsilon_0\simeq 0.6$ eV and $T_c\simeq 1741 K$. For a steady state coverage $c_0=0.9$, $\hat T_c\simeq 627 K$. A realistic value for the atomic diffusion coefficient around this temperature is $D_0\simeq 10^{-5}$ cm$^2$s$^{-1}$. As a result, for deposition rates in the range $\Gamma \simeq 10$ to $10^{3}$ $ s^{-1} $, in systems where $n=m=1$, $r=2$, $\theta_i\simeq 10^{-4} $ to $10^{-3} $ and $q_i \simeq 1.4\times 10^{7}$ to $4.47\times 10^{7}$ m$^{-1}$, which corresponds to a critical wavelength in the range $0.4$ to $0.14$ $\mu$m. In fact, for such experimental parameters, the critical wavelength behaves as $\lambda_i \simeq (\frac{l^2D_0}{\Gamma})^{1/4} $, or, for $l=4$ \AA\ and $\Gamma \simeq 10^{2}$ $ s^{-1}$ (which is in the range of typical values for deposition rates in e-beam evaporation, sputtering and CVD, $\lambda_i \simeq 4\times 10^{-4}(D_0)^{1/4} $ cm (for $D_0$ expressed in cm$^2$s$^{-1}$) and ranges from $40$ nm for $D_0= 10^{-8}$ cm$^2$s$^{-1}$ to $400$ nm for $D_0= 10^{-4}$ cm$^2$s$^{-1}$. This behavior is illustrated in figure \ref{figure3}.

On the other hand, close to threshold, $\frac{\Delta T_i}{T_i}=\frac{\nu^{2}}{\kappa} \simeq \frac{(1-2c_0)^{2}}{c_0(1-c_0)}$. For $c_0=0.9$, $\frac{\Delta T_i}{T_i}\simeq 7$ and hexagonal patterns may appear for $T<1.4 T_i$ while striped patterns are always unstable versus hexagonal ones. Hence, for $T<T_i$, hexagonal patterns should be the only stable ones. For $T_i<T<1.4 T_i$, hexagonal patterns and the uniform steady state are simultaneously stable, allowing the formation of localized hexagonal patterns in uniform background. In fact, it may be shown that stable stripes may only exist for $4\frac{(1-2c_0)^{2}}{c_0(1-c_0)}<1$ or $0.38<c_0<0.62$. Similar results results are obtained for systems where $m>r$, except that the uncovered steady state remains linearly stable and that stable stripes may only exist for $0.5<c_+<0.62$.

\begin{figure}
\includegraphics[width=4.5in]{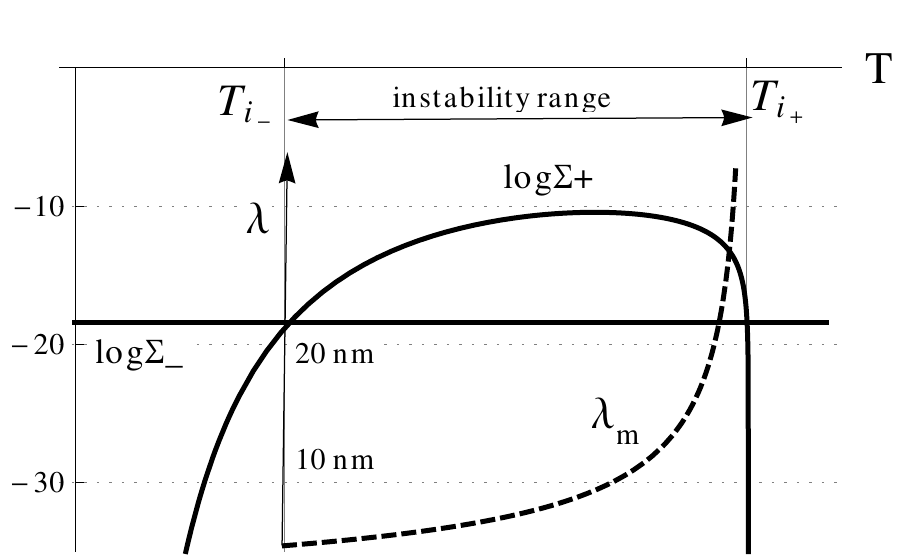}
 \caption{Schematic representation of the instability condition (\ref{lowTinstability2}) and the preferred wavelength, $\lambda_m$, as functions of temperature, for $D^*_0\simeq 3\times 10^{-2}$ cm$^2$s$^{-1}$, $\Omega \simeq 3760 K$, $l=5$ \AA\, $\epsilon_0\simeq 0.6$ eV, $c_+=0.9$ and $\alpha = 10^{2}$ $ s^{-1}$. } \label{figure4}
\end{figure}

If one considers thermally activated atomic transport, with the diffusion coefficient behaving as $D_0 = D^*_0 \exp -(\Omega /T)$, a realistic values for $D^*_0$ and $ \Omega $ may be $D^*_0\simeq 3\times 10^{-2}$ cm$^2$s$^{-1}$ and $\Omega \simeq 3760 K$, which corresponds to $D_0\simeq 1.4\times 10^{-4}$ cm$^2$s$^{-1}$ at $700 K$, $D_0\simeq 1.6\times 10^{-5}$ cm$^2$s$^{-1}$ at $500 K$ or $D_0\simeq 10^{-7}$ cm$^2$s$^{-1}$ at $300 K$. These values are consistent with the ones obtained, for example, through molecular dynamics simulations for surface diffusion coefficient on (100) $Al$ surfaces \cite{hanchen}. For $l=5$ \AA\, $\epsilon_0\simeq 0.6$ eV, $c_+=0.9$ and $\alpha = 10^{2}$ $ s^{-1}$, $T_c\simeq 1741 K$ and $\hat T_c\simeq 627 K$ . In this case $\Omega^*\simeq 6$ is not small, and the instability condition is illustrated in \ref{figure4}. It turns out that uniform coverage is unstable for $T_{i-}\simeq 147 K < T < T_{i-}\simeq 627(1-10^{-4})$. At the lower instability threshold, the preferred wavelength corresponds to 5 nm, while it corresponds to $0.438$ $\mu$m at the upper instability threshold. In fact, in the range $300 K < T < 600 K$ it goes from 6 to 20 nm.

\section{Conclusion.}\label{conclusions}

In this paper, the evolution of a growing monoatomic layer, deposited on a substrate, has been described by a dynamical model of the reaction-diffusion type. This dynamics combines reaction terms and nonlinear diffusion, and, close to the critical point of the order-disorder transition of the adsorbed layer, it corresponds to modified Cahn-Hilliard equations. Up to know, in this approach, the simplest deposition processes have been considered, which depend linearly on the atomic coverage of the growing film and may correspond to sputtering or laser assisted deposition. However, in chemical vapor deposition, the reaction part of the dynamics is more complicated and may also be nonlinear. Hence, when uniform deposited layers become unstable, the competition between nonlinearities arising from reaction or diffusion terms is expected to affect pattern formation, selection and stability in the deposited layer.

Different types of reaction dynamics have been considered. When the degree of nonlinearity is lower for adsorption than for desorption rates, deposition is dominant at low coverage and, favoring the initial growth of uniform layers which may be further destabilized by nonlinear diffusion. In this case, pattern formation phenomena should not qualitatively differ from the case described in \cite{walgraef_02b} where reaction terms are only due to linear adsorption and desorption processes. The effect of more complex reaction processes associated to chemical vapor deposition is only quantitative and modifies the existence and stability ranges of the different patterns, according to the relative importance of reaction and diffusion-induced nonlinearities. When the degree of nonlinearity is higher for adsorption than for desorption rates, desorption is dominant at low coverage and covered and uncovered domains may be simultaneously stable. According to temperature, covered domains may become unstable and develop spatial patterns . This may lead to a rich variety of structures which include arrays of patterned or uniform islands or dots in an uncovered background. At low temperatures, a Frank-Van der Merwe type of film growth with nanostructures is recovered. At high temperatures, however, localized island formation, which should initiate a Volmer-Weber type of film growth, can occur. Instability limits and stability domains for the different types patterns have been derived in the framework of weakly nonlinear analysis. The possible relevance of these results for realistic experimental data has been discussed and triggers the interest for further study, to confirm numerically the proposed pattern formation phenomena, on the one side, and to link them to experimental phenomena, on the other side.

\section*{Acknowledgments}
This work has been initiated in collaboration with Prof. N.M. Ghoniem. An association with the IFISC, at the University of the Balearic Islands (Spain) is also gratefully acknowledged.

\bibliography{ThinFilms}

\end{document}